\documentclass[aps,prd,twocolumn,a4paper]{revtex4-1}
\usepackage{ulem}
\usepackage{color}
\usepackage{graphicx}
\usepackage{hyperref}   
\usepackage{appendix}
\usepackage{epsfig}
\usepackage{epstopdf}
\usepackage{amsmath}
\usepackage{subfig}
\usepackage{comment}
\usepackage[dvipsnames]{xcolor}
\newcommand{\be}{\begin{equation}}
\newcommand{\ee}{\end{equation}}
\newcommand{\ba}{\begin{eqnarray}}
\newcommand{\ea}{\end{eqnarray}}

\begin{document}
\title{Impact of longitudinal bulk viscous effects to heavy quark transport in a strongly magnetized hot QCD medium}
\author{Manu Kurian}
\email{manu.kurian@iitgn.ac.in}
\affiliation{Indian Institute of Technology Gandhinagar, Gandhinagar-382355, Gujarat, India}
\author{Santosh K. Das}
\email{dsantoshphy@gmail.com}
\affiliation{School of Physical Sciences, Indian Institute of Technology Goa, Ponda-403401, Goa, India}
\author{Vinod Chandra}
\email{vchandra@iitgn.ac.in}
\affiliation{Indian Institute of Technology Gandhinagar, Gandhinagar-382355, Gujarat, India}

\begin{abstract}
The effects of longitudinal bulk viscous pressure on the heavy quark dynamics have been estimated in a 
strongly magnetized quark-gluon plasma within the Fokker-Planck approach. The bulk viscous modification 
to the momentum distribution of bulk degrees of freedom has been obtained in the presence of a magnetic 
field while incorporating the realistic equation of state of the hot magnetized QCD medium. As the magnetic 
field breaks the isotropy of the medium, the analysis is done along the directions longitudinal and transverse 
to the field. The longitudinal bulk viscous contribution is seen to have sizable effects in the heavy quark momentum 
diffusion in the magnetized medium. The dependence of higher Landau levels and the equation of state on the 
viscous correction to the heavy quark transport has been explored in the analysis.
\end{abstract}
\maketitle

 \section{Introduction}
 The very recent Large Hadron Collider (LHC) measurements provide a first sign of the existence of a strong 
electromagnetic field in the heavy-ion collision by measuring the directed flow $v_1$ for charged hadrons and 
$D/\bar{D}^0$ mesons for Pb$+$Pb collision at $\sqrt{s}_{NN}=5.02$ TeV~\cite{Acharya:2019ijj}. Several 
investigations have been done in the analysis of $v_1$ of hadrons with heavy quarks (HQs) incorporating 
the effects of a strong electromagnetic field~\cite{Das:2016cwd, Chatterjee:2018lsx, Coci:2019nyr}. The LHC 
measurements, together with the observation of Relativistic Heavy-Ion Collider (RHIC)~\cite{Adam:2019wnk}, 
indicate that the strong electromagnetic field created at the early stages of the collision affects the dynamics 
of the HQs. The HQs are mostly created in the very initial stages of the heavy-ion collision and travel through 
the deconfined hot nuclear matter-Quark Gluon Plasma (QGP). The HQs witness the entire QGP evolution as 
the thermalization time of HQ is larger than the lifetime of the QGP created at RHIC and LHC. These aspects 
allow HQs to serve as a potential probe to characterize  the properties of the QGP in the heavy-ion 
collisions~\cite{Prino:2016cni,Andronic:2015wma,Rapp:2018qla,Aarts:2016hap,Dong:2019unq,Cao:2018ews,Das:2015ana}.

The study of hot nuclear matter with the strong magnetic field has gained huge attention over the last 
decade~\cite{Karmakar:2019tdp,Fukushima:2017lvb,Koothottil:2018akg,Dey:2019vkn,Hattori:2017qih}. 
The phenomenological aspects of the QGP in the strong magnetic field primarily lies in the direction of 
chiral magnetic effect~\cite{Fukushima:2008xe}, chiral vortical effect~\cite{Kharzeev:2015znc,Avkhadiev:2017fxj} 
and recently in the studies of global $\Lambda-$hyperon polarization in the RHIC~\cite{STAR:2017ckg}. 
The HQ dynamics in magnetized QGP and anisotropic medium is investigated in 
Refs.~\cite{Fukushima:2015wck,Kurian:2019nna,Finazzo:2016mhm,Kiritsis:2011ha,Machado:2013rta,
Li:2016bbh,Bonati:2015dka,Guo:2015nsa,Mamo:2016xco,Rajagopal:2015roa,Singh:2017nfa,Giataganas:2013zaa}. Those investigations 
assume the QGP as a thermalized static medium. Hence, it is an interesting task to extend the analysis 
to viscous magnetized QGP and take account the non-equilibrium contributions~\cite{Das:2012ck,Singh:2019cwi} to 
the HQ transport in the medium.

The shear viscous tensor was considered to be the dominant source of dissipation for a long 
time~\cite{Romatschke:2007mq}. However, there are theoretical indications to the enhanced bulk 
viscosity in the medium~\cite{Meyer:2007dy,Karsch:2007jc,Denicol:2009am}. The magnetic field breaks 
the isotropy of the system and gives rise to two bulk viscous coefficients (transverse and longitudinal) and 
five shear coefficients. The authors of the Refs.~\cite{Hattori:2017qih,Fukushima:2017lvb} suggests that the 
dimensional mismatch of the quarks and gluons in the strongly magnetized QGP may lead to $1\rightarrow 2$ 
processes in the medium. All the components of the shear and bulk viscous coefficients have not been 
explored fully yet. The results of Refs.~\cite{Hattori:2017qih,Kurian:2018qwb} revealed that the longitudinal 
component of the bulk viscosity that arises from the quark contribution is dominant in the strong magnetic field
and is larger than that in the absence of a magnetic field.  

In the current analysis, we have iteratively solved the effective Boltzmann equation in relaxation time 
approximation (RTA) to obtain non-equilibrium momentum distribution function encoding the mean-field 
contributions in the magnetized bulk viscous medium. We have subsequently derived the evolution equation 
of bulk viscous pressure and obtain the non-equilibrium effects to the screening mass in the magnetized medium. 
We have illustrated that the incorporation of bulk corrections affects the HQ transport in the magnetized 
medium and significantly suppresses the magnetic field induced anisotropy in the HQ momentum diffusion 
in the temperature regime not far from transition temperature.

\section{Heavy quark dynamics in magnetic medium}
HQs propagates through the thermal QGP medium while interacting with quarks and gluons via 
$2\leftrightarrow 2$ scattering and can be described as the Brownian 
motion~\cite{Svetitsky:1987gq,GolamMustafa:1997id}. As the dynamics of quarks and gluons are 
different in the presence of a strong magnetic field, the estimation of the quark and gluonic 
contribution to the HQ transport coefficients need to be done separately in the magnetized medium. 
The current focus is on the regime of the strong magnetic with $gT\ll \sqrt{\mid q_feB\mid}$, in which higher Landau level (HLL) contributions are significant.  
The random motion of HQ in the QGP medium can be described by the evolution of momentum 
distribution function $f_{HQ}$ within the framework of Fokker-Planck equation,
\begin{align}\label{1.1}
\dfrac{\partial f_{HQ}}{\partial t}=\dfrac{\partial}{\partial p_i}\bigg[A_i({\bf p})~f_{HQ}
+\dfrac{\partial}{\partial p_j}\big[B_{ij}({\bf p})~f_{HQ}\big]\bigg],
\end{align}
where $A_i$ and $B_{ij}$ respectively measure the HQ drag force and momentum diffusion 
in the medium and takes the forms as follows,
\begin{align}\label{1.2}
&A_i=<<\big({\bf p}-{\bf p}^{'}\big)_i>>, &&B_{ij}=<<\big({\bf p}-{\bf p}^{'}\big)_i\big({\bf p}-{\bf p}^{'}\big)_j>>,
\end{align}
for the process, $HQ(p)+l(k)\rightarrow HQ(p^{'})+l(k^{'})$, 
where $l$ stands for thermal particles in the magnetized medium, with $\mid\mathcal{M}_{HQ,g/q}\mid^2$ 
as the matrix element. The thermal average can be defined as,
\begin{align}\label{1.201}
<<&\mathcal{F}>>=\dfrac{1}{d_{HQ}}\dfrac{1}{2E_p}\int{\dfrac{d\Upsilon} {2E_k}}\int{\dfrac{d^3
{\bf p}^{'}}{(2\pi)^32E_{p^{'}}}}\int{\dfrac{d\Upsilon^{'}} {2E_{k^{'}}}}\nonumber\\
&\times\mid\mathcal{M}_{HQ,g/q}\mid^2(2\pi)^{n}\delta^{n}(p+k-p^{'}-k^{'})f_{g/q}({\bf k})\nonumber\\
&\times\Big(1+f_{g/q}({\bf k}^{'})\Big)\mathcal{F}.
\end{align}
The integration phase factor can be described from the dimensional reduction in the presence of 
strong magnetic field ${\bf{B}}=B\hat{z}$ and takes the form $d\Upsilon=\frac{d^3{\bf{k}}}{(2\pi)^3}$ for gluons and 
$d\Upsilon=\frac{\mid q_feB\mid}{2\pi}\sum_{l=0}^{\infty}\mu_l{\frac{dk_{z}}{2\pi}}$ for quarks in the 
magnetized medium, where $\mu_l=(2-\delta_{l0})$ is the spin degeneracy of the Landau levels. 
Here, $d_{HQ}$ is the  degeneracy of the HQ, $f_{g/q}$ is the momentum distribution in the thermal 
medium, and $n=(2, 4)$ for the quarks and gluons, respectively. Note that in the static limit 
${\bf p}\rightarrow 0$, $B_{ij}\rightarrow K\delta_{ij}$~\cite{Svetitsky:1987gq}, where $K$ is the 
diffusion coefficient of HQ. As the magnetic field induces a spatial anisotropy in the medium, 
one we need to consider the HQ dynamics parallel and perpendicular to the magnetic field.

\section{Bulk viscous corrections in magnetic field}

\subsubsection*{Near-equilibrium thermal distribution function}

Proper modelling of the system in the thermal equilibrium followed by the knowledge of the longitudinal bulk 
viscous part of the distribution function is needed for the effective description of the bulk viscous effects to 
HQ transport in a magnetized system.
For the system not very far from local thermal equilibrium, the momentum  distribution function has the form, 
\begin{equation}\label{1.4}
f_{g/q}=f^0_{g/q}+\delta f_{g/q},     
\end{equation}
with $\delta f_{g/q}/f^0_{g/q} \ll1$.
The effective fugacity quasiparticle model (EQPM) describes the thermal medium interactions via QCD 
equation of state (EoS) in terms of quark and gluon effective fugacities, $z_q$ and $z_g$ 
respectively~\cite{Chandra:2011en}. The equilibrium EQPM distribution functions in the presence 
of the magnetic field ${\bf{B}}=B\hat{z}$ have the forms,
\begin{align}\label{1.5}
&{f}^{0}_{q}\equiv{f}^{0l}_{q}=\dfrac{z_{q}\exp{(-\beta E^l_{k})}}{1+ z_{q}\exp{(-\beta E^l_{k})}},
&&{f}^0_{g}=\dfrac{z_{g}\exp{(-\beta E_{k})}}{1- z_{g}\exp{(-\beta E_{k})}}.
\end{align}
The quark in the strongly magnetized medium follow a $1+1-$dimensional dynamics and the energy 
dispersion can be described by Landau quantization, $ E_k\equiv E^l_{k}=\sqrt{k_{z}^{2}+m_f^{2}+2l\mid q_feB\mid}$,  
where $l=0,1,2,..$ is the order of the Landau levels of the quark of mass $m_f$ and charge $q_fe$. 
The effective fugacity parameter modifies the single particle dispersion relation as,
$\omega^l_{q}=E^l_{k}+\delta\omega_q$,
and $\omega_{g}=E_k+\delta\omega_g$,
where the modified part of the non-trivial dispersion relation, $\delta\omega_{q/g}=T^{2}\partial_{T} \ln(z_{q/g})$, 
can be interpreted as the quasiparticle collective excitations in the medium. We consider the recent $(2+1)$ 
flavor lattice QCD EoS in the current analysis~\cite{Cheng:2007jq}. 

Transport coefficients are essential inputs to describe the non-equilibrium correction to the distribution function. 
In one dimensional system, both shear and bulk viscosities lead to similar hydrodynamical evolution as both 
viscosities corresponds to the same space-time gradient in the Navier-Stokes limit~\cite{Paquet:2019npk}. 
The longitudinal bulk viscous pressure in the strongly magnetized medium takes the form,~\cite{Kurian:2018qwb},
\begin{align}\label{1.14}
&\Pi_{\|}=-\sum_{l=0}^{\infty}\sum_f\mu_l\dfrac{\mid {q_f}eB\mid}{\pi}{N_c}
\int_{-\infty}^{\infty}{\dfrac{d\bar{k}_{z}}{2\pi\omega^{l}_q}
{\Delta_{\|}}_{\mu\nu}}\bar{k_{\|}}^{\mu}\bar{k_{\|}}^{\nu}
 \delta f^{l}_q\nonumber\\
&-\sum_{l=0}^{\infty}\sum_f\delta\omega_q\mu_l\dfrac{\mid {q_f}eB\mid}{\pi}{N_c}\int_{-\infty}^{\infty}{\dfrac{d\bar{k}_{z}}
{2\pi{\omega^l_{q}}}{\Delta_{\|}}_{\mu\nu}}\dfrac{\bar{k_{\|}}^{\mu}\bar{k_{\|}}^{\nu}}
{E^l_{k}}  \delta f^{l}_q,
\end{align}
where $\bar{k}^{\mu}$ is the covariant form of (dressed) quasiquark four-momentum and satisfy
$\bar{k_{\|}}^{\mu}=k_{\|}^{\mu}+\delta\omega_q u^{\mu}$,
with $\bar{k_{\|}}^{\mu}=(\omega^l_q,0,0,\bar{k}_z)$. Here, the longitudinal projection operator 
takes the form $\Delta_{\|}^{\mu\nu}\equiv g_{\|}^{\mu\nu}-u^{\mu}u^{\nu}$,
with $g_{\|}^{\mu\nu}=$ diag $(1,0,0,-1)$. The non-equilibrium part of the distribution function $\delta f^{l}_q$ 
can be obtained from the effective relativistic Boltzmann equation. The Boltzmann equation takes the form in the RTA as,
\begin{equation}\label{1.16}
\dfrac{1}{\omega^l_{_{q}}}\bar{k_{\|}}^{\mu}\partial_{\mu}f^{0l}_q(x,\bar{k}_{z})+F_q^{\mu}{\partial^{(k)}_{\mu} f_q^{0l}}=
-\dfrac{\delta f_q^{l}}{\tau_{R}},
\end{equation}
where $\tau_R$ is the thermal relaxation time and $F_q^{\mu}=-\partial_{\nu}(\delta\omega_q u^{\nu}u^{\mu})$ is 
the mean field force term that arises from the conservation laws of particle density and energy 
momentum~\cite{Mitra:2018akk}. We employ Chapman-Enskog like iterative expansion to solve the 
Boltzmann equation to describe $\delta f^{l}_q$ and has the following form for the first order correction to distribution function,
\begin{align}\label{1.17}
\!\!\delta f_q &= \tau_R\bigg[ \bar{k_{\|}}^\gamma\partial_\gamma \beta \!+\! \frac{\bar{k_{\|}}^\gamma}
{u\!\cdot\!\bar{k_{\|}}} \! \beta\, \bar{k_{\|}}^\phi \partial_\gamma u_\phi \! \!-\! \beta\theta_{\|}\,\delta\omega_q \bigg]f^{0l}_q\Tilde{f}^{0l}_q,
\end{align}
where $\Tilde{f}^{0l}_q=1-{f}^{0l}$ and $\theta_{\|}\equiv \partial_zu^z$ denotes the longitudinal expansion 
parameter of the magnetized system. Invoking the energy-momentum conservation laws, one can 
obtain $\dot{\beta}=\chi_{\beta}\theta_{\|}$, where $\chi_{\beta}/\beta\equiv c_s^2
=\frac{\partial P_{\|}}{\partial \varepsilon_{\|}}$ is the square of speed of sound in the longitudinal 
direction of the magnetized medium. By substituting Eq.~(\ref{1.17}) to Eq.~(\ref{1.14}) and assuming 
that $\tau_R$ is independent of four-momenta, we obtain first-order equation to the longitudinal bulk pressure as,
\begin{equation}\label{1.18}
\Pi_{\|}=-\tau_R\beta_{\Pi_{\|}}\theta_{\|},
\end{equation}
with the longitudinal bulk viscous coefficient $\beta_{\Pi_{\|}}$ as 
\begin{align}\label{1.19}
  \beta_{\Pi_{\|}}=&~\beta\bigg[ 
\frac{\chi_\beta}{\beta}\bigg({J}_{q~31}^{(0)}+\delta\omega_q{L}_{q~31}^{(0)}\bigg)\nonumber\\ 
&+{3}\bigg({J}_{q~42}^{(1)}+\delta\omega_q{L}_{q~42}^{(1)}\bigg)-\delta\omega_q{J}_{q~21}^{(0)}\bigg].  
\end{align}
The thermodynamic integrals ${J}_{k~nq}^{(r)}$ and ${L}_{k~nq}^{(r)}$ employed in the analysis 
are presented in the Appendix in terms of modified Bessel function of second kind. 
The magnetic field dependence of the longitudinal bulk pressure is incorporated through $\beta_{\Pi_{\|}}$. By comparing the above equation of longitudinal bulk viscous pressure with the relativistic Naiver-Stokes equation, we obtain the relaxation time for the bulk viscous expansion as $\tau_{\Pi}\equiv \tau_R =\frac{\zeta}{\beta_{\Pi_{\|}}}$, where $\zeta$ is the longitudinal bulk viscosity in the 
magnetized medium. The present analysis holds for an arbitrary process, and hence we choose different ranges of bulk viscosity.
Employing the bulk viscous evolution equation in Eq.~(\ref{1.18}), the longitudinal bulk viscous correction 
to the distribution function takes the form,
\begin{align}\label{1.21}
    \delta f_q^{l~\text{bulk}}&=\dfrac{-\beta}{\beta_{\Pi_{\|}}(u.\bar{k}_{\|})}\Big[(u.\bar{k}_{\|})^2\dfrac{\chi_\beta}{\beta}-{k_z^2}-(u.\bar{k}_{\|})\delta\omega_q\Big] f^{0l}_q\Tilde{f}^{0l}_q\Pi_{\|}.
\end{align}
The analysis of non-equilibrium correction to distribution function has been done in Ref.~\cite{Jaiswal:2014isa} in the absence of a magnetic field for a system of finite mass and ideal EoS. The bulk viscous correction to the momentum distribution function will give non-equilibrium corrections 
to the screening mass, which in turn can affect the matrix element for the HQ-thermal particle scattering processes. 

\subsubsection*{Non-equilibrium correction to Debye screening}

The realization of the EQPM from the charge renormalization can be done by analyzing the screening 
mass in the medium. The Debye mass in the magnetized QGP can be defined in terms of the EQPM 
distribution function and has the following form,
\begin{equation}\label{1.22}
{\bar{m}}_{D}^{2}=-4\pi\alpha_{s}\int{d\Upsilon
\dfrac{d}{d{\bf{k}}}(2N_{c}{f}_{g}+ 2N_{f}{f}^l_{q})},
\end{equation}
where $\alpha_{s}(T)$ is the running coupling constant. Note that the current analysis is on the strong field 
limit and the dominant quark contribution reduce to the form,  
\begin{align}\label{1.23}
\bar{m}_{D}^{2}=\dfrac{4\alpha_{s}}{T}\sum_f\dfrac{\mid q_feB\mid}{\pi}\sum_{l=0}^{\infty}\mu_l
\int_{0}^{\infty}{dk_z{f}^l_q(1-{f}^l_q)}.
\end{align}

The Debye mass in the viscous medium can be defined in the leading order as,
\begin{equation}\label{1.24}
 \bar{m}_{D}^{2}=m_{D}^{2}+\delta m_{D}^{2},  
\end{equation}
where $m_{D}^{2}$ denotes the screening mass in the medium at equilibrium, and $\delta m_{D}^{2}$ is the shift in Debye mass due to the longitudinal bulk viscous correction. Substituting Eq.~(\ref{1.4}) and Eq.~(\ref{1.5}) in Eq.~(\ref{1.23}), we have
\begin{align}\label{1.26}
{m}_{D}^{2}=\dfrac{4\pi\alpha_{s}}{3T}{J}^{(0)}_{q~10},
\end{align}
and 
\begin{align}\label{1.260}
\delta{m}_{D}^{2}=\frac{4\alpha_{s}}{T}\sum_f\frac{\mid q_feB\mid}{\pi}\sum_{l=0}^{\infty}\mu_l
\int_{0}^{\infty}{dk_z}F_q,
\end{align}
with $F_q=\delta f^l_q(1-2f^{0l}_q)$. Note that the bulk viscous correction to the Debye mass vanishes in the strong field limit in the massless case with ideal EoS ($z_{g/q}=1$). 

The effective running coupling constant within the EQPM, $\alpha_{eff}(T,z_q,z_g,\mid eB\mid)$ can be 
defined from $\bar{m}^{2}_{D}=\frac{\alpha_{eff}}{\alpha_s}{\bar{m}^{2}_{D~(z_k=1)}}$. The correction to 
the screening mass will reflect in the effective coupling and will act as essential dynamical input in 
the HQ dynamics in the QGP medium. 

\subsubsection*{Bulk viscous correction to the HQ drag transport}

The quark contribution to the momentum diffusion of HQ in the static limit in terms of momentum 
transfer can be defined from Eq.~(\ref{1.2}) and has the following form,
\begin{align}\label{1.27}
&K^{quark}_{\parallel}=\int{d^3{\bf q}\dfrac{d\Gamma}{d^3{\bf q}}q^2_z},
&&K^{quark}_{\perp}=\dfrac{1}{2}\int{d^3{\bf q}\dfrac{d\Gamma}{d^3{\bf q}}q_{\perp}^2},
\end{align}
where ${\bf q}={\bf p}-{\bf p}^{'}$ is the momentum transfer due to the interaction and 
\begin{align}\label{1.28}
\dfrac{d\Gamma}{d^3{\bf q}}=&\dfrac{1}{d_{HQ}}\dfrac{1}{2M_{HQ}}\dfrac{1}{(2\pi)^32
E_q}\int{\dfrac{d\Upsilon}{2E_k^l}}\int{\dfrac{d\Upsilon^{'}}{2E_{k^{'}}^l}}\mid\mathcal{M}_{HQ,q}\mid^2\nonumber\\
&\times(2\pi)^2\delta^2(p+k-p^{'}-k^{'})f^l_q(k_z)
\Big(1-f^l_q(k^{'}_z)\Big),
\end{align}
\begin{figure*}
 \centering
 \subfloat{\includegraphics[scale=0.34]{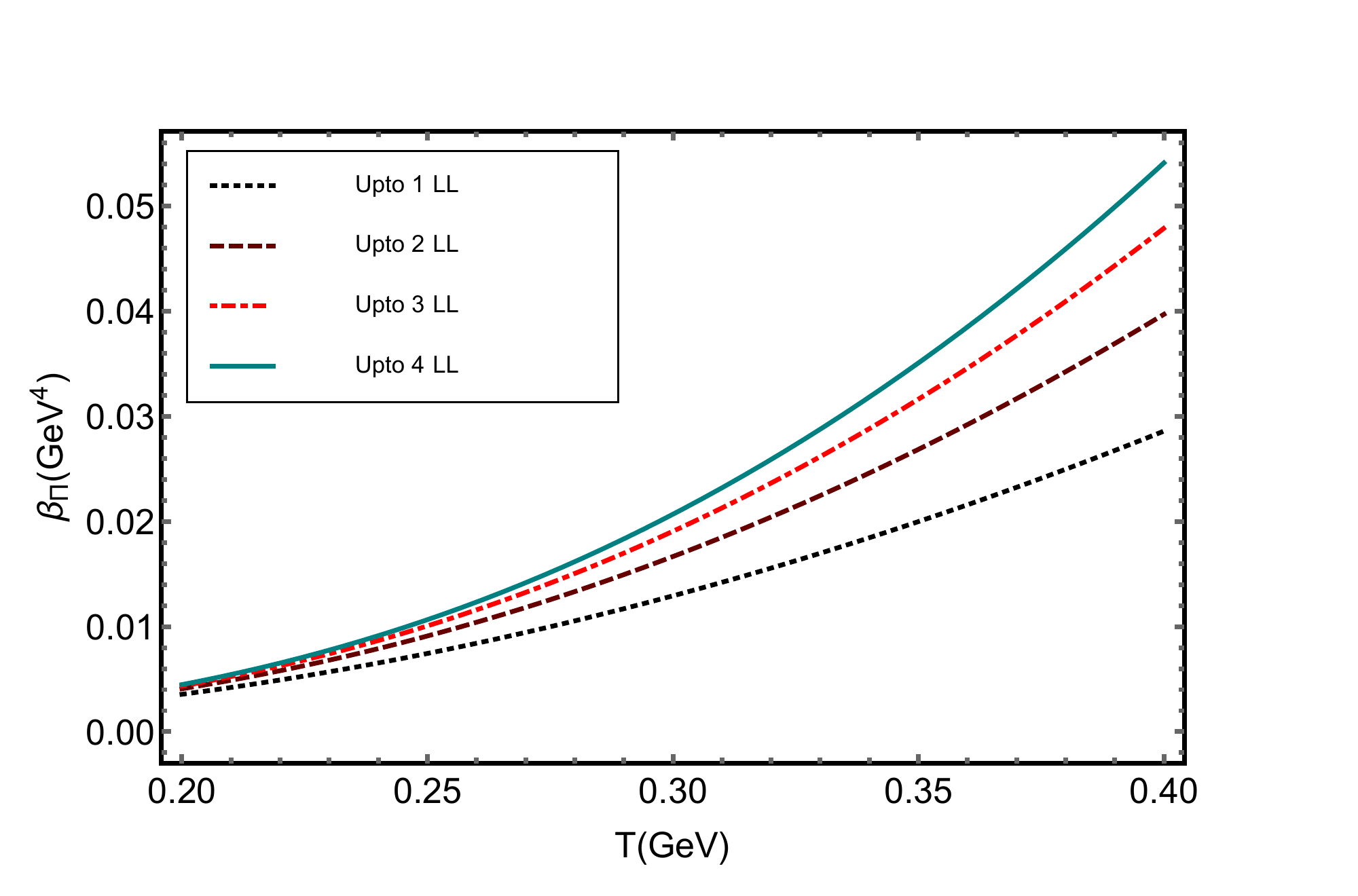}}
 \hspace{.6 cm}
 \subfloat{\includegraphics[scale=0.38]{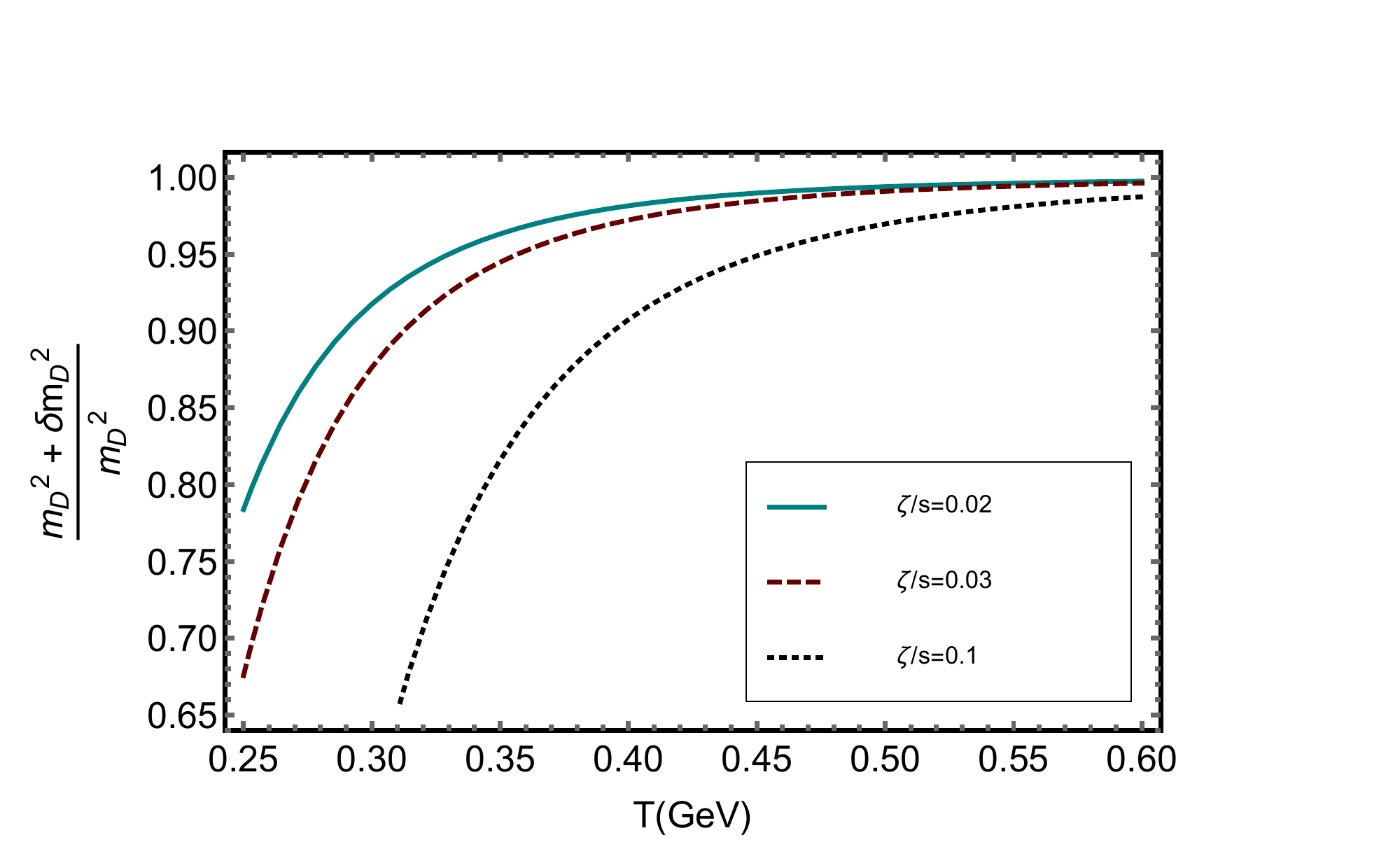}}
\caption{The temperature dependence of $\beta_{\Pi_{\|}}$ at $\mid eB\mid=15m_{\pi}^2$ (left panel). 
The effect of longitudinal bulk viscous effects to the screening mass at $\tau=0.3$ fm$^{-1}$(left panel).}
\label{f1}
\end{figure*} 
denotes the HQ-quark scattering rate per unit volume of momentum transfer with $M_{HQ}$ is 
the mass of HQ. The scattering rate can be defined from the retarded self-energy as follows~\cite{Fukushima:2015wck}, 
\begin{align}\label{1.29}
\dfrac{d\Gamma}{d^3{\bf q}}&= \dfrac{8\pi\alpha_{eff}C_R^{HQ}}{(2\pi)^3}\underset{\omega\rightarrow 0}
{\text{lim}}\dfrac{T}{\omega}\nonumber\\&\times\bigg[ \frac{{\text{Im}}\Pi_{R~\text{Fermion}}^{00}}
{(Q^2-{\text{Re}}\Pi_{R~\text{Fermion}}^{00})^2+({\text{Im}}\Pi_{R~\text{Fermion}}^{00})^2}\bigg],
\end{align}
where $C_R^{HQ}$ is the Casimir factor of the HQ. Here, $Q=(\omega, {\bf q})$ with $\omega$ is the 
energy transfer and static limit imposes the condition $\omega\rightarrow 0$. The leading order longitudinal bulk viscous corrections are incorporated through the distribution function while defining the retarded self-energy (via quark/antiquark loops) along with the equation of state effects in the medium. The definitions of the real and imaginary part of the self-energy are modified in the viscous medium, which, in turn, gives rise to a shift in Debye mass. A shift in the Debye mass in bulk viscous medium from the gluon-self energy in the absence of a magnetic field is studied in Ref.~\cite{Du:2016wdx}. 

The retarded quark propagator and the gluon polarization tensor at the one-loop order within the regime $gT\ll \sqrt{\mid eB\mid}$ by incorporating the effects of HLLs is discussed in Ref.~\cite{Fukushima:2017lvb}. The transverse components gluon self-energy vanishes (negligible) in the  leading order (one-loop order) and this results in the vanishing electrical conductivity within the regime of focus~\cite{Fukushima:2017lvb}.  The gluon self-energy takes the form $\Pi^{\mu\nu}_{R~\text{Fermion}}(Q)=i4\pi\alpha_sT_R<J_r^{\mu}(Q)J_a^{\nu}(-Q)>$, in the real-time Keldysh formalism in {\it ra} basis with $J^{\mu}=(J^0, j^i)$ as the current operator. In Refs.~\cite{Fukushima:2017lvb,Kurian:2018qwb}, the authors have showed that the HLLs give corrections to the longitudinal quark dynamics and modify the macroscopic quantities such as energy momentum tensor $T^{\mu\nu}$ and four-current $J^{\mu}$ of the medium  (transverse contributions are negligible), in the regime  $gT\ll \sqrt{\mid eB\mid}$. The transverse dynamics can be decouple from the longitudinal dynamics and we have, $\Pi^{\mu\nu}_{R~\text{Fermion}}(Q)=\pi \bar{s}(q_{\perp}) \Pi^{\mu\nu}_{R~1+1}(\omega, q_z)$. The estimation of gluon self-energy and the screening mass with all Landau levels is described in the Ref.~\cite{Ghosh:2018xhh}. It is important to emphasize that we consider the Landau approximation as most of the scattering processes are soft (small momentum transfer $\sim gT$). The  $1+1-$dimensional retarded self-energy $\Pi^{\mu\nu}_{R~1+1}(\omega, q_z)$  is defined as 
$\Pi^{\mu\nu}_{R~1+1}(\omega, q_z)=i<J_r^{\mu}(q_{\parallel})J_a^{\nu}(-q_{\parallel})>$, 
with $<J_r^{\mu}(q_{\parallel})J_a^{\nu}(-q_{\parallel})>$ is the retarded 
 current-current correlator in $1+1-$dimension. The quantity $\bar{s}(q_{\perp})$ defines from the real part of the retarded self energy in the bulk viscous medium, $\bar{s}(q_{\perp})\equiv {s}(q_{\perp})+\delta{s}(q_{\perp})=\underset{\omega\rightarrow 0}{\text{Re}} \Pi^{00}_{R~\text{Fermion}}(q)$. In the limit of LLL approximation and ideal EoS, $\bar{s}(q_{\perp})$ reduces back to the form in the Ref.~\cite{Fukushima:2015wck}. The real and imaginary part of the retarded self-energy takes the form in the viscous medium,
\begin{align}\label{1.30}
&{\text{Re}} \Pi^{00}_{R~\text{Fermion}}(\omega, {\bf q})=-\dfrac{ q_z^2}{q^2_{\parallel}}\bar{s}(q_{\perp}),\\
&{\text{Im}} \Pi^{00}_{R~\text{Fermion}}(\omega, {\bf q})=\dfrac{\pi \omega}{2}\bar{s}(q_{\perp})
\big[\delta(\omega-q_z)+\delta(\omega+q_z)\big],\label{1.300}
\end{align}
where quantity $\bar{s}(q_{\perp})$ can be defined as,
\begin{align}\label{1.31}
\bar{s}(q_{\perp})=&4\pi\alpha_{s}\sum_f\frac{1}{T}\frac{\mid q_feB\mid}{\pi^2}\sum_{l=0}^{\infty}\mu_l
\int_{0}^{\infty}{dk_z}\nonumber\\&\times{f}^l_q(1-{f}^l_q)\exp{\Big(\dfrac{-q_{\perp}^2}{2\mid q_feB\mid}\Big)},\\
&\equiv {s}(q_{\perp})+\delta {s}(q_{\perp}).
\end{align}
Note that $\bar{s}(q_{\perp}=0)$ denotes the leading order contribution (quark part) to the Debye 
screening mass in the strongly magnetized viscous medium. Hence, the viscous correction to 
the ${s}(q_{\perp}=0)$ can be defined as $\delta {s}(q_{\perp})=\delta m_D^2 
\exp{\big(\frac{-q_{\perp}^2}{2\mid q_feB\mid}\big)}$.

The delta function in Eq.~(\ref{1.300}) can be understood from the $1+1-$dimensional constrained motion of the quarks in the regime $gT\ll \sqrt{\mid eB\mid}$.  Since our focus is on soft momentum transfer limit ($\sim gT$), the interactions do not change the Landau levels of the quarks in the static limit as the energy gap associated with adjacent Landau levels  $\Delta\epsilon\sim \sqrt{q_feB}$ is much greater than the energy transfer in the process, $\Delta\epsilon \gg \omega$, in the current regime of focus. Therefore, the energy-momentum transfer of the quark $(\Delta E, \Delta k_z)=(\omega, q_z)$ satisfy $\omega\simeq\pm q_z$ by neglecting the term $\mathcal{O}(1/eB^2)$ as in the case of LLL  in Ref.~\cite{Fukushima:2015wck}.
The static limit $\omega\rightarrow 0$ further imposes the vanishing longitudinal momentum transfer, denoted as $\delta(q_z)$. It is important to note that the Eq.~(\ref{1.30}) and Eq.~(\ref{1.300}) will not hold for the non-static limit and also in the weak/moderate magnetic field beyond the regime $gT\ll \sqrt{\mid eB\mid}$ where transverse components are non-negligible. This is beyond the scope of the current analysis. Substituting Eqs.~(\ref{1.29})-(\ref{1.31}) in the Eq.~(\ref{1.27}) and defining $x=\frac{q^2_{\perp}}{2\mid eB\mid}$, we have
\begin{align}\label{1.32}
\bar{K}_{\perp}^{quark}=K_{\perp}^{quark}+\delta K_{\perp}^{quark} \end{align}
where the equilibrium part can be defined as, 
\begin{align}\label{1.33}
K_{\perp}^{quark}=&4~T~\alpha_{eff}~\alpha_{s}N_cC_R^{HQ}~\dfrac{{\mid eB\mid}}{2\pi}\nonumber\\
&\times\int_{0}^{\infty}{dx\dfrac{{x}~\mathcal{N}(T,x)}{\Big(x+2\dfrac{\alpha_{s}}{\pi}\mathcal{N}(T,x)\Big)^2}},
\end{align}
and the leading order longitudinal bulk viscous correction to the transverse component takes the form,
\begin{align}\label{1.35}
\delta K_{\perp}^{quark}=&4~T~\alpha_{eff}~\alpha_{s}N_cC_R^{HQ}~\dfrac{{\mid eB\mid}}{2\pi}\nonumber\\
&\times\int_{0}^{\infty}{dx\dfrac{{x^2}~\delta \mathcal{N}(T,x)}{\Big(x+2\dfrac{\alpha_{s}}{\pi} \mathcal{N}(T,x)\Big)^3}}.
\end{align}
The quantity $\mathcal{N}(T,x)$ and $\delta \mathcal{N}(T,x)$ takes the following form respectively,
\begin{align}\label{1.36}
&\mathcal{N}=\dfrac{1}{T}\sum_f\mid q_f\mid e^{-\frac{x}{\mid q_f\mid}}\sum_{l=0}^{\infty}\mu_l
\int_{0}^{\infty}{dk_z{f}^{0l}_q\Tilde{f}^{0l}_q},\\
&\delta \mathcal{N}=\dfrac{1}{T}\sum_f\mid q_f\mid e^{-\frac{x}{\mid q_f\mid}}\sum_{l=0}^{\infty}\mu_l
\int_{0}^{\infty}{dk_z F_q}.
\end{align}
The vanishing longitudinal component of the quark contribution in the static limit $\omega\rightarrow 0$ 
is well explored in the Ref.~\cite{Fukushima:2015wck,Kurian:2019nna} and can be understood from 
the Eqs.~(\ref{1.27}),~(\ref{1.29}) and~(\ref{1.300}).
\begin{figure*}
 \centering
 \subfloat{\includegraphics[scale=0.36]{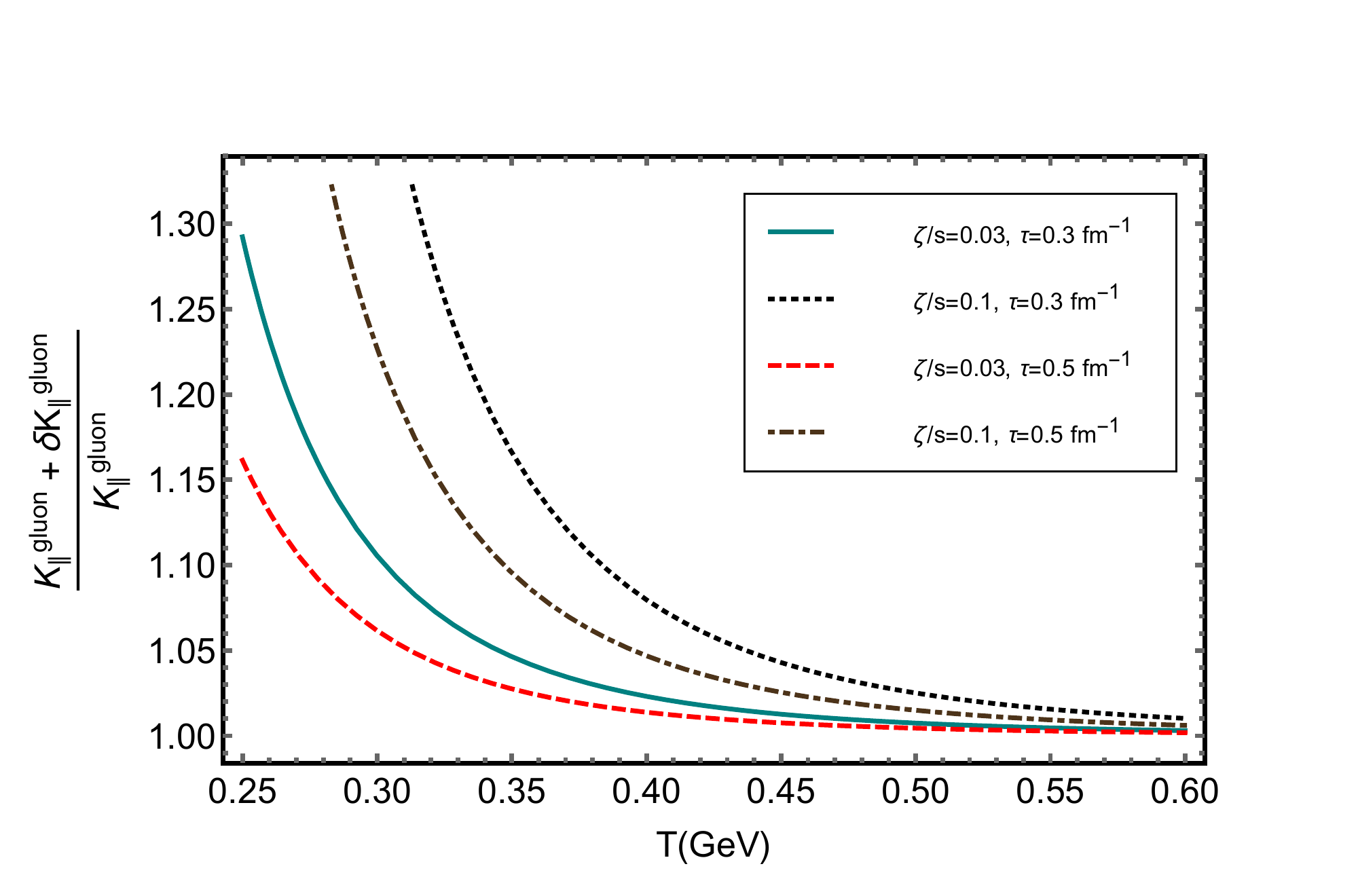}}
 \hspace{.65cm}
 \subfloat{\includegraphics[scale=0.36]{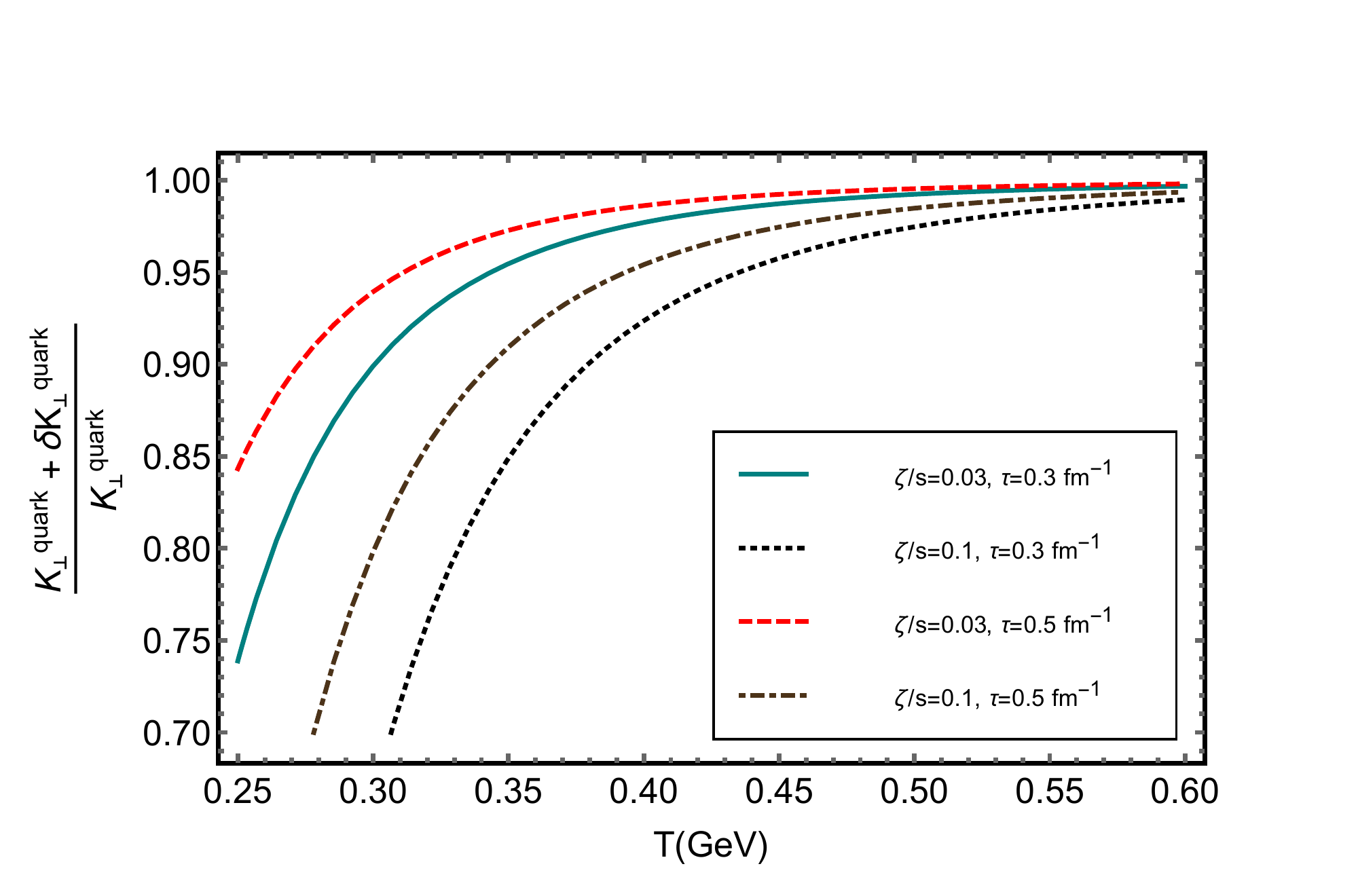}}
\caption{The longitudinal bulk viscous correction to $K^{gluon}_{\parallel}$ and $K^{quark}_{\perp}$ 
with $l=4$ at $\mid eB\mid=15m_{\pi}^2$.}
\label{f2}
\end{figure*} 
\begin{figure*}
 \centering
 \subfloat{\includegraphics[scale=0.35]{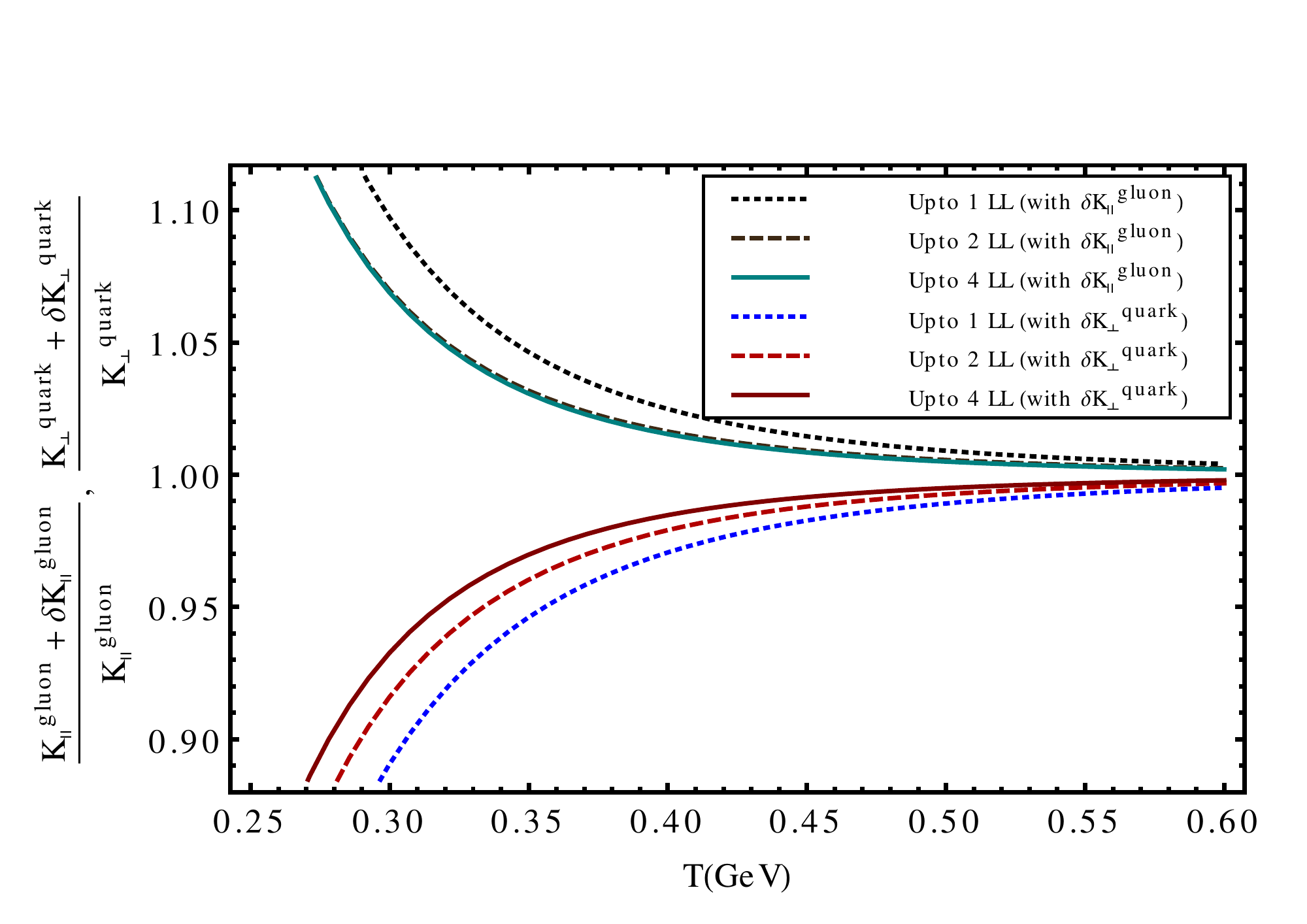}}
 \hspace{1.5cm}
 \subfloat{\includegraphics[scale=0.35]{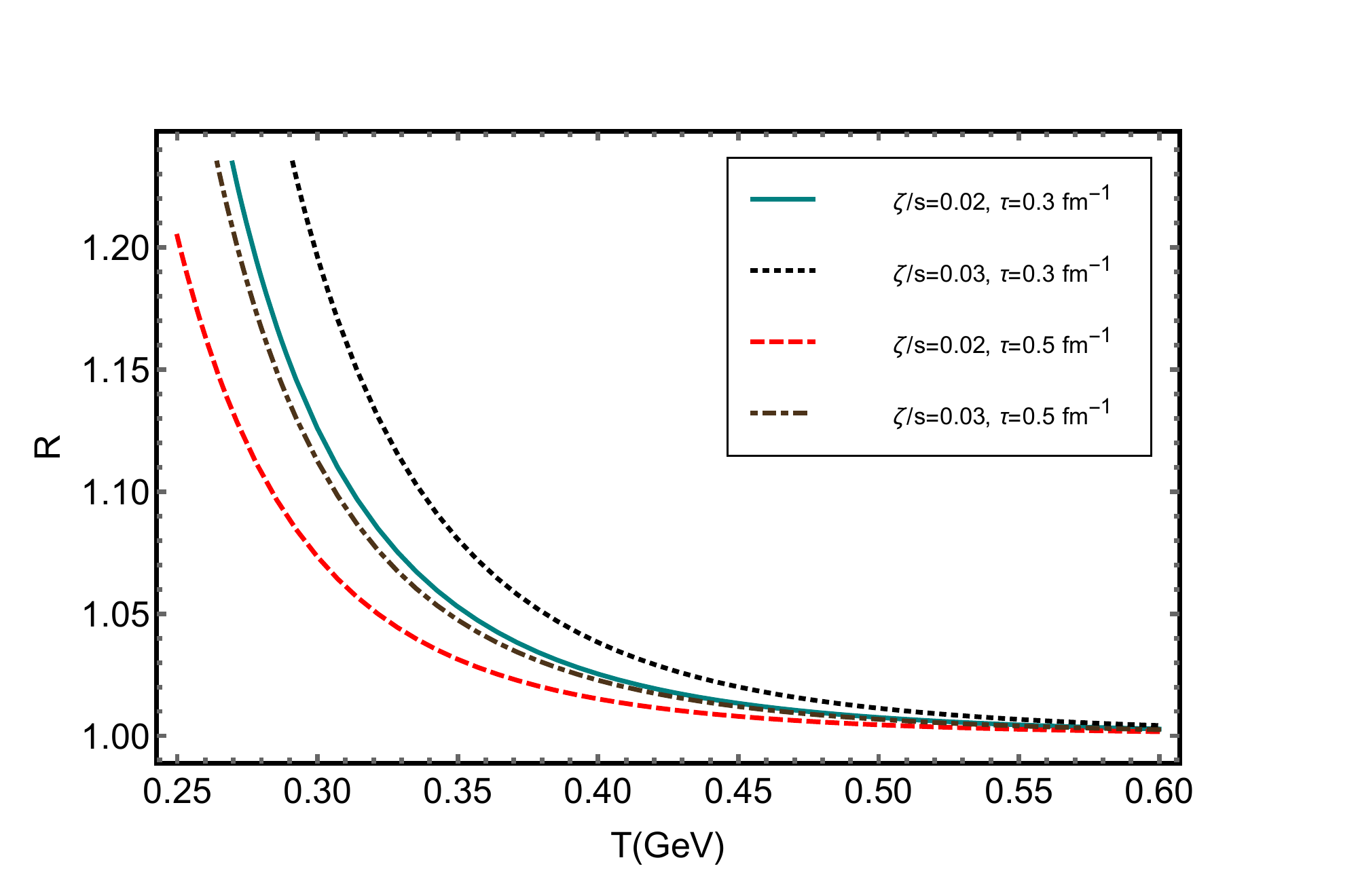}}
\caption{The HLLs effect in the longitudinal bulk viscous correction to the HQ diffusion at 
$\mid eB\mid=15m_{\pi}^2$ (left panel). The effect of bulk viscous corrections to the magnetic 
field induced HQ momentum anisotropy (right panel).}
\label{f3}
\end{figure*} 

The bulk viscous correction to the gluonic contribution to the HQ momentum diffusion is primarily 
incorporated through the screening mass while defining the HQ-gluon matrix element. The matrix 
element for the $2\leftrightarrow 2$ HQ-gluon scattering process in the static limit is investigated in 
the Ref.~\cite{Fukushima:2015wck}. Following the same prescriptions as in Ref.~\cite{Kurian:2019nna} 
to describe the longitudinal component of the gluonic contribution in the viscous medium, we obtain
\begin{equation}\label{1.37}
\bar{K}^{gluon}_{\parallel}=K^{gluon}_{\parallel}+\delta K^{gluon}_{\parallel},    
\end{equation}
\begin{align}\label{1.38}
K^{gluon}_{\parallel}&=\dfrac{4}{{3}\pi}\alpha_{eff}^2N_cC_R^{HQ}\dfrac{1}{d_{HQ}}
\int_{0}^{\infty}{dq}~\dfrac{q^2}{\big(q^2+s(q_{\perp})\big)^2}\nonumber\\
&\times\int_{q/2}^{\infty}{d\mid{\bf k}\mid}\mid{\bf k}\mid^2\Big[1
+\Big(1-\dfrac{q^2}{2\mid{\bf k}\mid^2}\Big)^2\Big] f^0_g\Tilde{f}^0_g,
\end{align}
\begin{align}\label{1.39}
\delta K^{gluon}_{\parallel}&=-\dfrac{4}{{3}\pi}\alpha_{eff}^2N_cC_R^{HQ}\dfrac{1}{d_{HQ}}
\int_{0}^{\infty}{dq}~\dfrac{q^2\delta s(q_{\perp}) }{\big(q^2+s(q_{\perp})\big)^3}\nonumber\\
&\times\int_{q/2}^{\infty}{d\mid{\bf k}\mid}\mid{\bf k}\mid^2\Big[1+\Big(1-\dfrac{q^2}
{2\mid{\bf k}\mid^2}\Big)^2\Big] f^0_g\Tilde{f}^0_g,
\end{align}
where $\Tilde{f}^0_g=1+f^0_g$. 
The gluonic contribution to the HQ momentum diffusion in the magnetized medium is isotropic up to 
the leading order of $m_D/T$, whereas the quark contributions are highly anisotropic in nature. 
The anisotropy of the HQ momentum diffusion can be quantified by the ratio 
$\frac{K_{\parallel}}{K_{\perp}}$ in which $K_{\parallel}$ and $K_{\perp}$ denotes the 
total contribution (both quark and gluonic) to the longitudinal and transverse components 
of the HQ momentum diffusion in the QGP medium.  The effect bulk viscosity to the 
anisotropy can be described from 
$R=\frac{{\bar{K}_{\parallel}}/{\bar{K}_{\perp}}}{{K_{\parallel}}/{K_{\perp}}}$.

\section{Discussions}
We initiate the discussion with the temperature dependence of first order longitudinal bulk viscous 
coefficient of the strongly magnetized QGP medium as plotted in Fig.~\ref{f1} (left panel). We observe that the first order bulk viscous coefficient increases with temperature. 
This observation is consistent with the results for the Boltzmann system~\cite{Bhalerao:2013aha} in 
which $\beta_{\Pi}\approx P$ with $P$ as the pressure of the QGP. The effects of HLLs 
are more visible in the higher temperature range as the effect is suppressed by the factor 
$e^{-\frac{\sqrt{q_feB}}{T}}$. 
The longitudinal bulk viscous corrections to the Debye screening mass is depicted in Fig.~\ref{f1} (right panel). 
To quantify the effects of the bulk viscous effects, we choose the expansion parameter $\theta_{\|}=1/\tau$, 
where $\tau$ is the proper time parameter. 
The Debye mass is sensitive to the viscous corrections, and we observe that the longitudinal bulk 
viscous contribution reduces the screening in the magnetized medium. This observation is in 
line with that of the Ref.~\cite{Du:2016wdx}.  

The temperature dependence of the longitudinal bulk viscous correction to the gluonic contribution 
$\bar{K}^{gluon}_{\parallel}$ in the longitudinal direction at $\mid eB\mid=15m_{\pi}^2$ is plotted in Fig.~\ref{f2}. 
The longitudinal bulk viscous correction enhances the gluonic contribution to the HQ diffusion coefficient.
We have estimated the quark contribution in the perpendicular direction in the viscous strongly 
magnetized QGP. Bulk viscous correction reduces the perpendicular quark contribution to the HQ 
momentum diffusion, and the effects are more visible in the lower temperature regimes. 
The quark-HQ scattering rate per unit volume of momentum transfer is proportional to the term $\frac{\bar{s}(q_{\perp})}{(q^2+\bar{s}(q_{\perp}))^2}$. In the leading order (of coupling constant), we have $\frac{\bar{s}(q_{\perp})}{(q^2+\bar{s}(q_{\perp}))^2}\simeq\frac{{s}(q_{\perp})}{(q^2+{s}(q_{\perp}))^2}+\frac{q^2\delta{s}(q_{\perp})}{(q^2+\bar{s}(q_{\perp}))^3}$ and the scattering rate gets suppressed, as $\delta s(q_{\perp})$ quantifies the decrease in the screening mass in the medium. Thus, the suppression of the quark-HQ scattering rate due to the off-equilibrium effects from the scattering matrix and the quark distribution function leads to the suppression of the quark contribution of HQ momentum diffusion in the bulk viscous medium.

The HLLs effects to the bulk viscous corrections of the HQ momentum diffusion are more significant 
in the temperature regime near to the transition temperature. We observe in Fig.\ref{f3} that the 
anisotropy of the momentum diffusion of the HQ in the viscous QGP decreases in comparison with 
the thermally equilibrated system. The longitudinal bulk viscous effects are more pronounced in the 
lower temperature regime near transition temperature, and asymptotically the ratio $R$ approaches unity.

\section{Conclusion and Outlook}
To summarize, we have estimated the longitudinal bulk viscous evolution equation and obtained the 
bulk viscous correction to the thermal distribution in the magnetic medium by solving the effective 
Boltzmann equation within the EQPM. We have illustrated that the bulk viscous contribution reduces 
the Debye screening mass in the magnetized hot QCD medium. We have studied the HQ momentum 
diffusion in the magnetized viscous medium. The main observation is that the bulk viscous corrections 
suppress the quark contribution to the HQ momentum diffusion, whereas the gluonic contributions to 
the HQ diffusion gets enhanced. This, in turn, affects the magnetic field induced 
anisotropy in the HQ momentum diffusion in the medium. We have further demonstrated the effects 
of HLLs on the bulk viscous corrections to the HQ diffusion in the magnetized medium.

The recent LHC observation on heavy mesons directed flow $v_1$ 
give the indications of the strong electromagnetic field. 
However, recent calculations~\cite{Das:2016cwd, Chatterjee:2018lsx, Coci:2019nyr}  on the heavy meson $v_1$ due to
the electromagnetic field, within the Langevin dynamics,
ignore the impact of the magnetic field and viscosity on HQ drag and diffusion
coefficients. The anisotropic HQ transport coefficients in the strongly 
magnetized viscous medium may affect the heavy meson directed flow measured 
both at RHIC and LHC energies. Heavy meson elliptic flow is another experimentally 
measured observable which can be affected by this anisotropic HQ transport 
coefficients.  The HQ transport coefficients presented in the manuscript will be used in 
Langevin dynamics  as inputs parameters to study HQ observables in the presence of 
the strong magnetic field.  The effect of $1\rightarrow 2$ processes and $2\rightarrow 2$ 
processes in the expanding medium to the HQ experimental observables is another interesting direction to explore.
We intend to extend the analysis to these aspects of the hot QCD medium 
in a forthcoming article.

\section{Acknowledgments}
V.C. acknowledge SERB for the Early 
Career Research Award (ECRA/2016), and the DST, Govt. of India for INSPIRE-Faculty Fellowship 
(IFA-13/PH-55). S.K.D. acknowledge the support by the National Science Foundation of China 
(Grants No.11805087 and No. 11875153).

\appendix

 \section{Thermodynamic integrals in the magnetic field}\label{A}

Defining, $y^2_l=\frac{1}{T^2}(m_f^2+2l\mid q_feB\mid)$ and
$K_{i,n}(sy_l)=\int_{0}^{\infty}{\frac{d\theta}{(\cosh{\theta})^n}\exp{(-sy_l\cosh{\theta})}}$,
and following the prescriptions in Ref.~\cite{Bhadury:2019xdf}, the ${J}^{(r)}_{q~mn}$ and ${L}^{(r)}_{q~mn}$ 
integrals in the presence of magnetic field takes the following forms,
\begin{flalign}
    {J}^{(0)}_{q~10} =&T\sum_fN_c\frac{\mid q_feB\mid}{\pi^2}\sum_{l=0}^{\infty}\mu_ly_l\sum_{s=1}^{\infty} s(-1)^{s-1}z_q^sK_1(sy_l),
\end{flalign}
\begin{flalign}
    {J}^{(0)}_{q~31} =&-T^3\sum_fN_c\frac{\mid q_feB\mid}{4\pi^2}\sum_{l=0}^{\infty}\mu_ly_l^3\sum_{s=1}^{\infty} s(-1)^{s-1}z_q^s\nonumber\\&\times\bigg[K_3(sy_l)-K_1(sy_l)\bigg],
\end{flalign}
\begin{flalign}
    {L}^{(0)}_{q~31} =&-T^2\sum_fN_c\frac{\mid q_feB\mid}{2\pi^2}\sum_{l=0}^{\infty}\mu_ly_l^2\sum_{s=1}^{\infty} s(-1)^{s-1}z_q^s\nonumber\\&\times\bigg[K_2(sy_l)-K_0(sy_l)\bigg],
\end{flalign}
\begin{flalign}
    {L}^{(1)}_{q~42} =&T^2\sum_fN_c\frac{\mid q_feB\mid}{2\pi^2}\sum_{l=0}^{\infty}\mu_ly_l^2\sum_{s=1}^{\infty} s(-1)^{s-1}z_q^s\nonumber\\&\times\bigg[K_2(sy_l)-3K_0(sy_l)+2K_{i,2}(sy_l)\bigg],
\end{flalign}
\begin{flalign}
    {J}^{(1)}_{q~42} =&T^3\sum_fN_c\frac{\mid q_feB\mid}{4\pi^2}\sum_{l=0}^{\infty}\mu_ly_l^3\sum_{s=1}^{\infty} s(-1)^{s-1}z_q^s\nonumber\\&\times\bigg[K_3(sy_l)-5K_1(sy_l)+4K_{i,1}(sy_l)\bigg]\nonumber\\
    &-\delta\omega_qT^2\sum_fN_c\frac{\mid q_feB\mid}{\pi^2}\sum_{l=0}^{\infty}\mu_ly_l^2\sum_{s=1}^{\infty} s(-1)^{s-1}z_q^s\nonumber\\&\times\bigg[K_2(sy_l)-3K_0(sy_l)+2K_{i,2}(sy_l)\bigg],
\end{flalign}
\begin{flalign}
    {J}^{(0)}_{q~21} =&-T^2\sum_fN_c\frac{\mid q_feB\mid}{2\pi^2}\sum_{l=0}^{\infty}\mu_ly_l^2\sum_{s=1}^{\infty} s(-1)^{s-1}z_q^s\nonumber\\&\times\bigg[K_2(sy_l)-K_0(sy_l)\bigg]\nonumber\\
    &+\delta\omega_qT\sum_fN_c\frac{\mid q_feB\mid}{\pi^2}\sum_{l=0}^{\infty}\mu_ly_l\sum_{s=1}^{\infty} s(-1)^{s-1}z_q^s\nonumber\\&\times\bigg[K_1(sy_l)-K_{i,2}(sy_l)\bigg].
\end{flalign}
Note that the thermodynamic integrals in terms of the modified Bessel function of the second kind are for the general case, and in the massless limit, the integral with $K_n(sy_0)$ reduced to PolyLog functions as discussed in Ref.~\cite{Bhadury:2019xdf} in detail.



{}

\end{document}